\documentclass[conference]{IEEEtran}
\usepackage{graphicx}
\usepackage[cmex10]{amsmath}
\usepackage{algorithmic}
\usepackage{array}
\usepackage[tight,footnotesize]{subfigure}
\usepackage{fixltx2e}
\usepackage{url}
\usepackage{graphics}
\usepackage{amssymb}

\ifCLASSOPTIONcompsoc
  \usepackage[nocompress]{cite}
\else
  \usepackage{cite}
\fi

\begin{document}

\title{Experimental study and pratical realization of a reconciliation method for quantum key distribution system}
\author{\IEEEauthorblockN{N. Benletaief, H. Rezig *Members IEEE, A. Bouallegue *Members IEEE}
\IEEEauthorblockA{Communication System laboratory Sys'Com \\National Engineering School of Tunis\\ BP 37, 1002 Tunis  Belv\'ed\`ere, Tunisia\\
Emails: benletaief.nedra@gmail.com, houria.rezig@enit.rnu.tn, ammar.bouallegue@enit.rnu.tn}}
\maketitle

\begin{abstract}
This paper investigates a reconciliation method in order to establish an errorless secret key
in a QKD protocol. Classical key distribution protocols are no longer unconditionally secure because computational complexity of mathematical problems forced hardships. In this context, QKD protocols offer a highest level of security because they are based on the quantum laws of physics. But, the protocol performances can be lowered by multiples errors. It appears clearly that reconciliation should be performed in such a situation in order to remove the errors as for the legitimate partners. The proposed method accomplishes reconciliation by using QTC  in the special problem of side-information source coding ("Slepian-Wolf" coding model).
Our theoretical hypothesis are sustained by experimental results that confirm the advantage of our method in resolving reconciliation problem compared to a recent related work. Indeed, the integration of our method  generates an important progess in security and a large decrease of the QBER. The gain is obtained with a reasonable complexity increase. Also, the novelty of our work is that it tested the reconciliation method on a real photonic system under VPItransmissionMaker.
\end{abstract}

\section{Introduction}
\vspace{-2pt}
Society aspires to more security of remote information exchanges. To ensure safety under these conditions, it is necessary to encrypt the data using a secret key. But the exchange of this key remains an essential step for many systems. Undeniably, quantum computers are expected to highly perform in the next few years endangering  public key cryptosystems  safety. In fact, their safety is often based on some computational assumption which involves large and randomly generated keys. One solution is to apply QKD.
\\Our work aimes at finding reconciliation solution to the QKD protocol. While a brief problem statement is presented in  section 2, an overview of previous related works is drown in section 3. In Section 4,  we describe our proposed solution to the problem of reconciliation. Section 5 will provide comparative discussion with related work in terms of security and correcting ability. Also, we present selected experimental results of VPItransmissionMaker.
\vspace{-6pt}
\section{Problem statement}
\vspace{-2pt}
Cryptography relies on transforms involving a shared secret key to transform the communicated plaintext. The key distribution aims to share this theoretic secret key from two legitimate parties Alice and Bob thereof in the presence of an eavesdropper Eve. Classical information which Alice and Bob want to exchange are carried on quantum signals. For instance, a classical bit is encoded into the polarization or the phase of a photon. The first quantum cryptography protocol was proposed in 1984 by Bennett hence its name the BB84 protocol \cite{Bennett}. The procedure of BB84 protocol will be illustrated below.
\begin{enumerate}
\item[(i)] Alice takes a string of bits and basis randomly in order to prepare quantum states. Then, she sends them to Bob using the quantum channel. On receiving the state, Bob measures them and informs Alice through public channels of the used basis. After that, Alice informs Bob, which bits were used in the correct basis. At this point, whenever they use the same basis, they have a great probability to have correlated results (this probability is influenced by the presence of eavesdropping and the channel imperfection). They omit the bits corresponding to incompatible choice of basis.  The remaining correlated random variables are termed sifted keys.
\item[(ii)] The channel parameters are estimated by Alice and Bob from announced bits such as the joint probability distribution among the two partners and the eavesdropper.
\item[(iii)] In this step, in order to improve BB84 performances by conversation over the public channel, we resold to reconciliation.
\item[(iv)] Eve may have partial information of the key after reconciliation process. Thus, amplification is needed to lower down Eve's information by making  bits almost statistically independent of all the information gathered by Eve.
\end{enumerate}
 This paper focus in the step of reconciliation in order to correct errors in the sifted keys.  The BB84 protocol is provided with an unconditionally proofs of security in \cite{prof3,prof4} that result of physics' quantum laws. But, errors caused by imperfect channel and eavesdropping can reduce the protocol performances. Designing and validation an powerful reconciliation method  is a crucial task for QKD. Next, we consider some related works before announcing our proposed method.
\vspace{-6pt}
\section{Related works}
\vspace{-2pt}
Several methods of information reconciliation was investigated in the literature. The first one to be considered is Binary algorithm proposed by Bennett et $al$.~\cite{Binary} which applied an interactive error correction based on binary searching . It has the advantage to be simple and easy.  A stronger ability of error correction method appeared later and it is called Cascade by Brassard et $al$.~\cite{Cascade}. Reducing the information loss using an interactive protocol such as Cascade and Binaary is theoretically feasible. However, this leads in practice to a remarkably overload in communication and to a highly limitation of the effective rate of the key. Also, the communication's time of Cascade  algorithm dependends  on the  key's length where it depends only on the error rate for the winnow algorithm \cite{Winnow}. Winnow uses the syndrome from a Hamming code as a property of forward error correcting to correct the error in a block with different parity. The performance of error detection and correction is limited with Hamming code that still far from the theoretical limit. Later a modified one way error reconciliation which employs a mixed Hamming code concatenation scheme is proposed by \cite{performham}. The improved protocol not only maximizes the key generation rate but also promotes correction capability.  Besides these method of reconciliation, it exits methods based on errors correcting codes. In this direction, the QLDPC codes have recently shown the same ability of error correcting as Cascade and present the advantage to boost the protocol security \cite{rate}, \cite{rateadap1}, \cite{rateadap2}, \cite{rateadap3}.

\vspace{-6pt}
\section{Proposed method for QDK reconciliation}
\vspace{-2pt}
Recently, QTC have shown to be good candidates for this reconciliation application \cite{nedra_jqis}. At the opposite of \cite{nedra_jqis} which deals with parallel QTC, we propose to use in our work serial QTC to accomplish reconciliation according to the analysis model "Slepian-Wolf". As advocated by the laws of quantum mechanics cloning of qubits is impractical in reality. Thus, parallel QTC are also infeasible.
We start with a brief description of "slepian-wolf coding".
\subsection{Slepian-Wolf coding \cite{SWcoding}}
\begin{figure}[h]
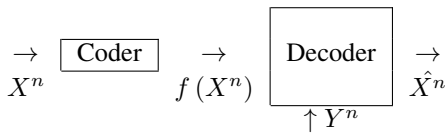

\begin{centering}
\begin{tabular}{ccc|c|cc}
\cline{4-4}
& & & & & \tabularnewline
\cline{2-2}
\multicolumn{1}{c|}{$\rightarrow$} & \multicolumn{1}{c|}{Coder} & $\rightarrow$ & Decoder & $\rightarrow$ & \tabularnewline
\cline{2-2}
$X^{n}$ & & $f\left(X^{n}\right)$ & & $\hat{X^{n}}$ & \tabularnewline
\cline{4-4}
& & \multicolumn{1}{c}{} & \multicolumn{1}{c}{$\uparrow Y^{n}$} & & \tabularnewline
\end{tabular}
\par\end{centering}
\caption{ "Slepian-Wolf" Model. \label{Fig:slepian-wolf}}
\end{figure}

Reconciliation with error correcting codes is based on what is commonly called the "Slepian-Wolf" coding. As shown in the figure Fig.\ref{Fig:slepian-wolf}, this technique consists of an agreement between Alice and bob on a rate coding of the code $R$ that can correct on average more than ($\frac{pn}{2}$) errors, where $p$ refers to the probability of errors. Alice then communicates to Bob the syndrome $X^{n}$ in a public way (or the reverse). Bob thanks to his knowledge of $Y^{n}$ and the syndrome of $X^{n}$ can found $\hat{X^{n}}$ as the closest code word $Y^{n}$ having the good syndrome. The encoder and the decoder are assumed to have a good estimation of the joint probability distribution $P_ {X^{n}Y^{n}}$.
\\Reconciliation differs from error correction coding by the fact Alice lost the advantage of choosing what she sends. Indeed, messages can not be limited to codewords of a given code. But, in order to consider the encoding formalism:
\begin{itemize}
\item $Y^{n}$ can be seen as a received code word with one exception that the syndrome $X^{n}$ is no longer 0 but $HX^{n}$ where $H$ is the parity matrix of the code.
\item The parity of a control node in the decoding "Slepian-Wolf" is determined from $HX^{n}$
\end{itemize}
\subsection{Quantum Turbo codes(QTC)}
The QTC have been less studied than the QLDPC codes because of quantum convolutional codes. In the quantum case, these codes can not verify two properties (recursiveness and noncatastrophic error propagation) necessary for a good performance of the QTC. In fact, they generate a unbounded minimum distance. Also, they generate iterative decoding algorithm that reduces the noise each iteration converging towards a good estimation of the error. This did not prevent authors in to give solutions in \cite{turboexemple,quantum}.

\subsubsection{Coding principle for quantum Turbo codes}
\begin{figure}[h]
\centering
\includegraphics[width=4cm,height=1.5cm]{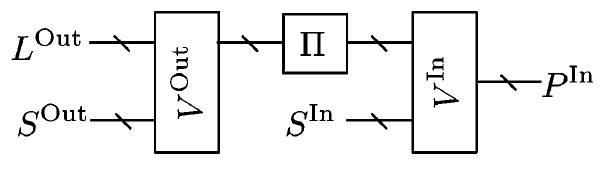}
\caption{Quantum Serial Turbo code.\label{Fig:turbo}}
\end{figure}

The codes are designed by using two component codes. The information is interleaved before entering the second code. The codes that constitute the QTC are two identical non recursive and nonsystematic convolutional codes (see Fig.~\ref{Fig:turbo}).
\\  Both the use of an iterative principle in decoding and the presence of an interleaver shapes the efficiency of the QTC.
\subsubsection{Decoding principle for quantum Turbo codes}
\begin{figure}[h]
\centering
\includegraphics[width=7cm,height=4cm]{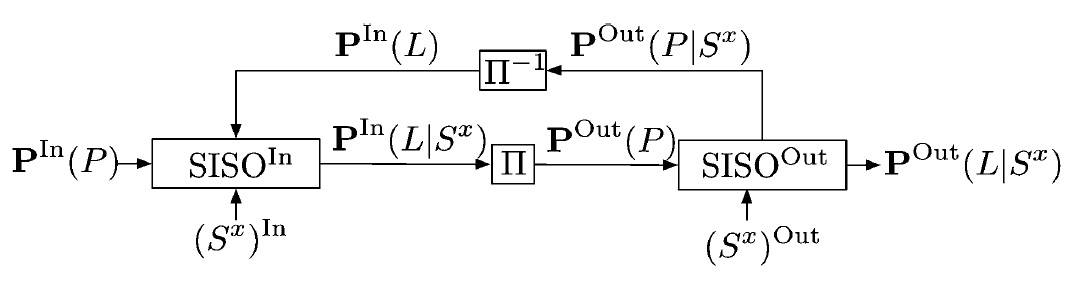}
\caption{Quantum Serial Turbo decoder.\label{Fig:turbodec}}
\end{figure}

In the quantum case, it exist different versions for decoding quantum stabilizer codes.  Generally, the turbo decoder consists of two, or more, maximum likelihood decoders (see Fig.~\ref{Fig:turbodec}). In our work, we choose to optimize the algorithm proposed by \cite{LDPCthes}.
We used the procedure for decoding QTC operated on memoryless Pauli channels (Errors are distributed according to a product distribution). The decoding algorithm is an adaptation to the quantum case of the usual SISO algorithm used for decoding serial Turbo of codes \cite{SISO}. But, the decoding algorithm in the quantum case differs from the standard version in number of points.
\begin{itemize}
\item As measuring the quantum state may disturb it, we consider Pauli's error which affected the quantum state instead of the state of qubits.
\item The maximum likelihood decoding is to identify the most likely error knowing the syndrome.
\item Memory is related the Pauli's error affecting qubits and don't reflect the encoded state.
\end{itemize}
\vspace{-6pt}
\section{Results}
\vspace{-2pt}
In what follows,  we discuss the experimental performances of our reconciliation method operating on binary variables which are applied to discrete variable QKD.
\subsection{Experimental conditions}
We have implemented and experimented our reconciliation method on the case of one special type of eavesdropping strategy (Intercept and Resend) and a special memoryless channel (depolarizing channel).
\begin{itemize}
 \item For intercept and Resend attack, Eve try to guess the quantum states sent by Alice before sending replacement states to Bob. Errors can occur in the key Alice and Bob share if Alice and Eve don't choose the same basis.
 \item The depolarizing channel on $n$ qubits of error probability $p$ is a special case of memoryless channel where errors are chosen independently.

\begin{equation}
\begin{cases}
\begin{array}{c}
P\left(E=\varepsilon\right)=\frac{p}{3}\qquad\forall\varepsilon\in\left\{ X,Y,Z\right\} \\
P\left(E=I\right)=1-p
\end{array}\end{cases}
\end{equation}
\end{itemize}
For more generalities we take into account two possible constructions for QLDPC codes and QTC which are summarized in the Table ~\ref{Tab:condexp}. QLDPC codes are chosen to be regular stabilzer codes and QTC are noncatastrophic and nonrecursive codes.
\begin{table}[h]

\caption{Code's setting. \label{Tab:condexp}}
{\scriptsize }%
\centerline{\footnotesize
\begin{tabular}{|c|c|}
\hline
{\scriptsize }%
\begin{tabular}{c}
QLDPC \tabularnewline
\tabularnewline
\end{tabular} & {\small }%
\begin{tabular}{|c|c|c|c|}
\hline
 & \multicolumn{1}{c||}{{\small $n$}} & {\small $k$} & {\small $R$}\tabularnewline
\hline
{\small $LDPC1(6,12)$} & {\small 3600} & {\small 1800} & {\small $\frac{1}{2}$}\tabularnewline
\hline
{\small $LDPC2(4,8)$} & {\small 8736} & {\small 4370} & {\small $\frac{1}{2}$}\tabularnewline
\hline
\end{tabular}\tabularnewline
\hline
\begin{tabular}{c}
QTC \tabularnewline
\tabularnewline
\end{tabular} & {\small }%
\begin{tabular}{|c|c|c|c|c|c|}
\hline
 & {\small $n$} & {\small $k$} & {\small $m$} & {\small $\mathtt{R}conv$} & {\small $R$}\tabularnewline
\hline
\hline
{\small $Turbo1$} & {\small 3} & {\small 1} & {\small 3} & {\small $\frac{1}{3}$} & {\small $\frac{1}{9}$}\tabularnewline
\cline{2-5}
 & {\small 3} & {\small 1} & {\small 3} & {\small $\frac{1}{3}$} & \tabularnewline
\hline
{\small $Turbo2$} & {\small 2} & {\small 1} & {\small 4} & {\small $\frac{1}{2}$} & {\small $\frac{1}{4}$}\tabularnewline
\cline{2-5}
 & {\small 2} & {\small 1} & {\small 4} & {\small $\frac{1}{2}$} & \tabularnewline
\hline
\end{tabular}\tabularnewline
\hline
\end{tabular}
}
\end{table}

In the next section, we  describe the results of our simulation by studying Quantum Bit Error Rate (QBER), secure information and complexity.
\subsection{Simulation results}
\subsubsection{Quantum Bit Error Rate}
\begin{figure}[h]
\centering
\includegraphics[width=10cm,height=7.5cm]{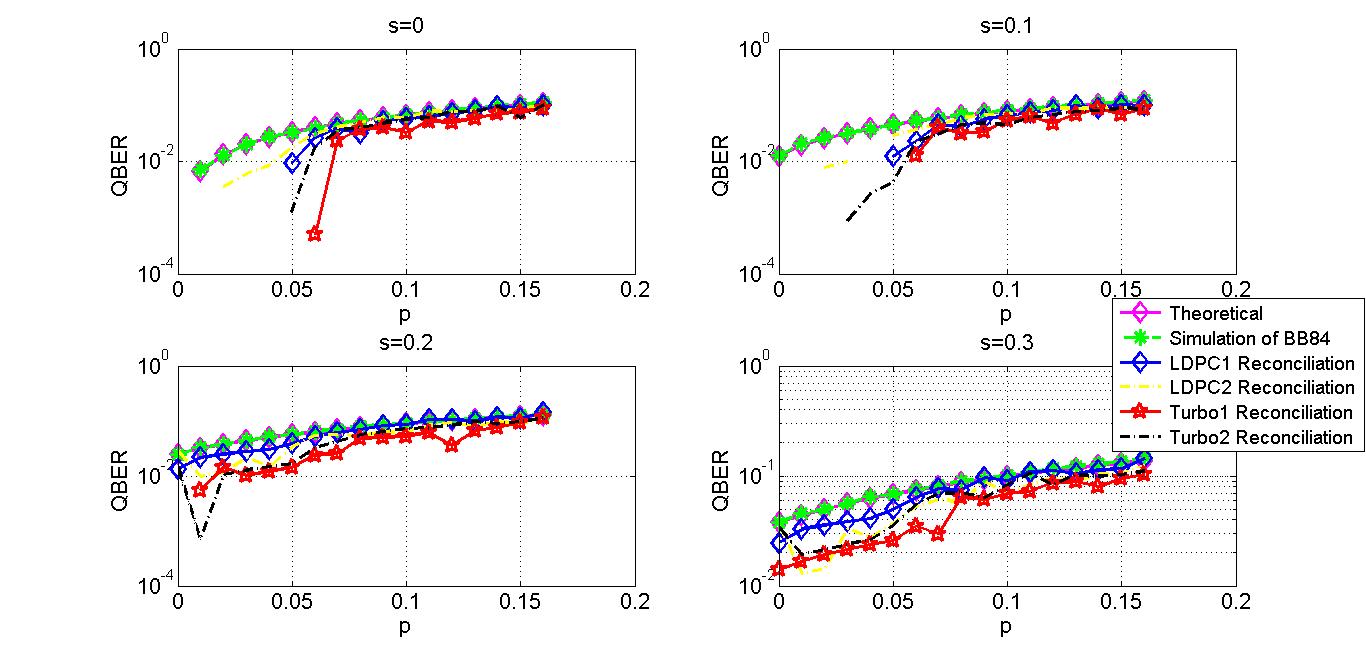}
\caption{QBER as function of $p$-values for Intercept and Resend attack for different $s$ values.\label{Fig:intercept1}}
\end{figure}

We represent in Fig.~\ref{Fig:intercept1} QBER as function of $p$-values for different probality $s$  that one qubit sent by Alice to Bob is intercepted. The performance of reconciliation method is calculated in terms of QBER compared to the one generated by the Intercept and Resend attack given by \cite{nedra_jqis}:
\begin{equation}
QBER=\frac{s}{4}.
\end{equation}
Fig.~\ref{Fig:intercept1} shows the good correction power of our approach of reconciliation. Indeed, on average, our method fixes more than 40 \% of errors caused by eavesdropping and depolarized channel.
\\However, as we can notice in Fig.~\ref{Fig:intercept1}, Turbo1 and Turbo2 have a "hashing bound" for a depolarization probability $p = 0.08$. But, this "hashing bound" remains higher than that of the LDPC1 and LDPC2 (P = 0.07). We can notice at this stage of our work that the results of our simulation platform are validated given the concordance observed between the bound of the simulation and the bound in the literature \cite{constrcutioncode}. At p = 0.08, the difference between the two competitive methods is no longer significant. Also, we note that the gain from the application of our method of reconciliation decreases as the evesdropping capacity increases. The reason for this decline is that the method is more suitable to correct the effects of depolarized channel by integrating its initialization parameters when decoding. However, the evesdropping model is not taken into account.
We can add at this stage of our study that despite the undoubted interest of the QTC in our case study, they remain strongly influenced by several parameters as the number of iterations to be considered and the type of interleaver to deploy. We start with the first one as it is essential to increase the speed of convergence and to achieve a minimum error rate.
\begin{figure}[h]
\centering
\includegraphics[width=8cm,height=6cm]{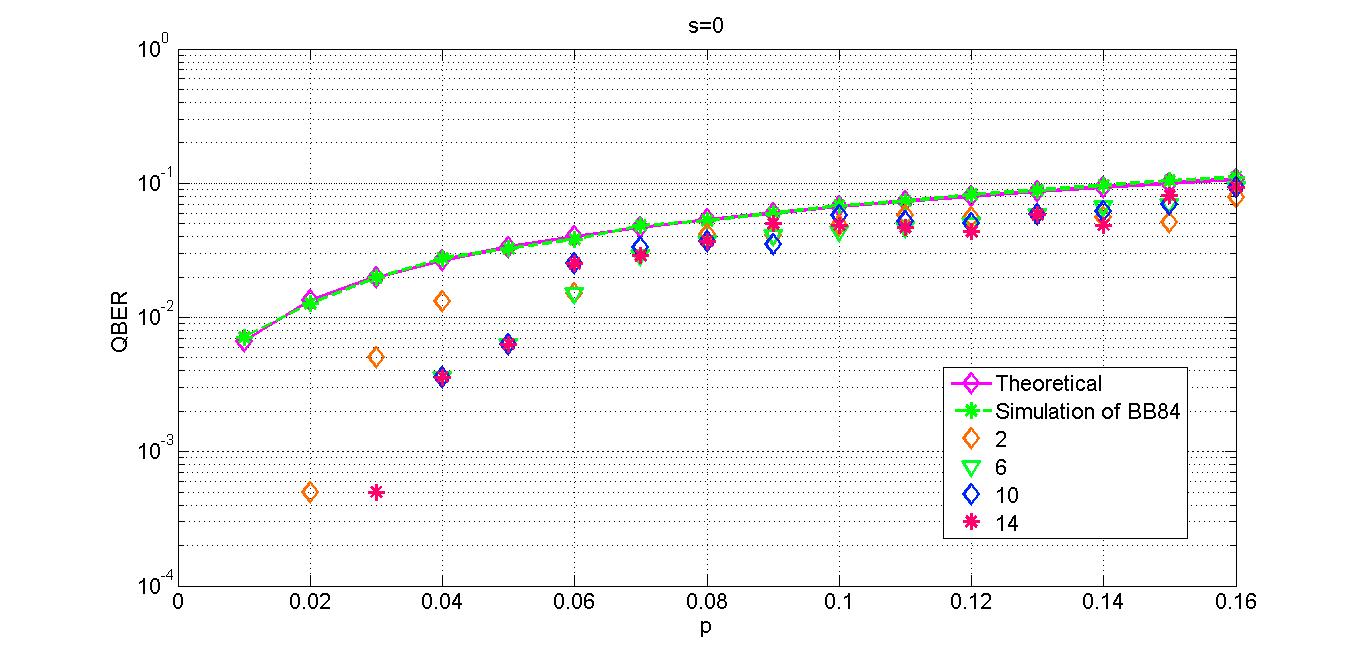}
\caption{QBER as function of $p$-values for different iteration values.\label{Fig:intercept3}}
\end{figure}
Fig.~\ref{Fig:intercept3} presents the QBER obtained after 2 to 14 iterations of decoding (Turbo1, $R=\frac {1} { 9}$). At low level of depolarization parameter $p$, we remark a change in slope of the curves according to the number of iterations. We also note that a number of iterations 10 is enough to have a good convergence.
Also, interleavers are part of the parameters to be taken into consideration. Initially , random interleaver interleaver1 is considered. Then, we adopt an interleaver that exploits Pauli's error interleaver2. This type of interleaving is characterized by knowing in advance the position of qubits after interleaving.
\begin{figure}[h]
\centering
\includegraphics[width=8cm,height=6cm]{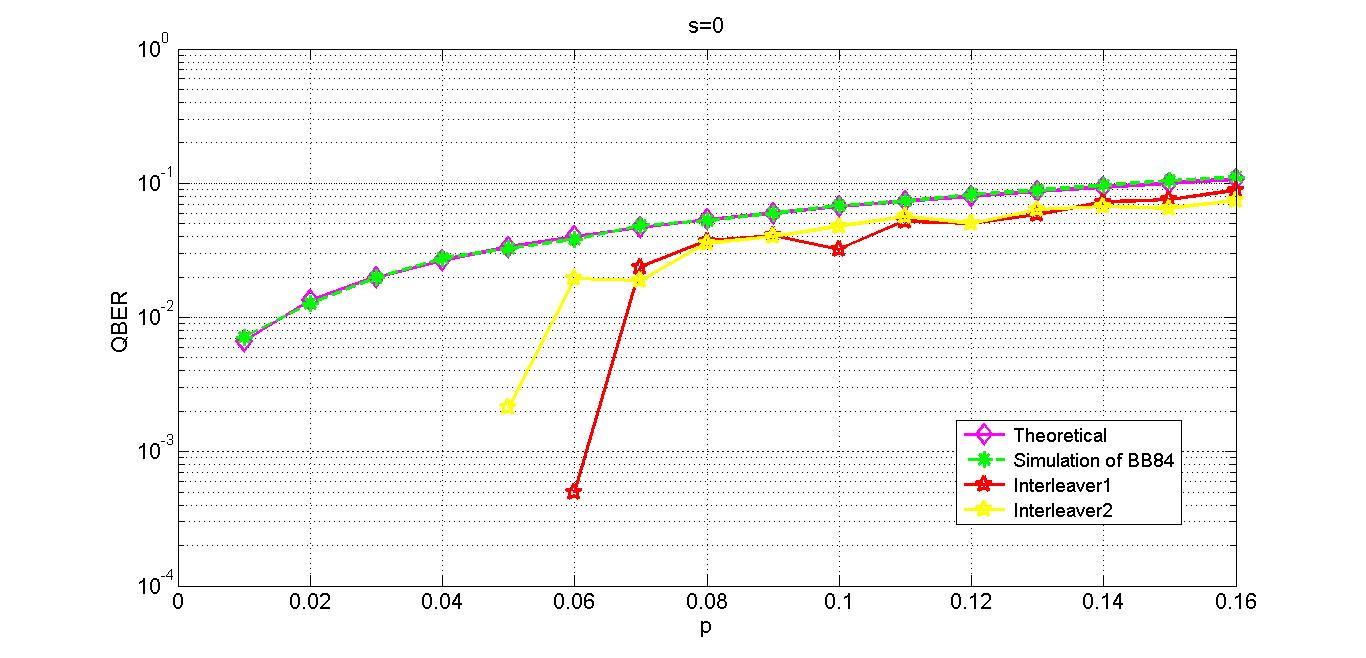}
\caption{QBER as function of $p$-values for different interleavers.\label{Fig:intercept2}}
\end{figure}
We note from Fig.~\ref{Fig:intercept2}, in the general case the random interleaving's performance are better than those of the modified interleaver.
\\The results given are due to several reasons that we will try to explain below.
\begin{itemize}
\item First, we note that one of the disadvantages of QLDPC codes LDPC1 (6; 12) and LDPC2 (4; 8)
is that they have many significant 4 cycles. As it is well known, cycles affects the iterative decoding's performance.
However, it should be emphasized that this is a special feature of all codes constructed as stabilizer formalism.
\item In addition, in the classical case, irregular LDPC codes outperform the regular ones. Unfortunately, in the quantum case, the generalisation of irregular LDPC codes has not yet given good performance. This explains the choice of regular codes in particular.
\item We can add that generally, the QLDPC codes constructed using the stabilizer code formalism tend to have a poor minimum distance.
\item In addition, we can notice that the correlation between Pauli's errors $X$ and $Z$ errors depends on depolarization parameter $p$. When $p$ increases, the correlation decreases. Therefore, performance decreases in this case.
\item Compared to QLDPC codes, QTC offer several advantages such as the liberty in designing the code in terms of length, rate, memory size and especially the choice of the interleaver. It is well known that interleavers are  the source of randomness and the major device establishing code's success. In the other hand, the iterative decoder in the case of QTC explicitly exploits the degeneracy of those codes. This property is important because the QTC such as QLDPC codes have lightweight stabilizers
and are therefore highly degenerate. If the character is degenerate, in theory capable of improving the transmission of quantum information, it appears in practice that degrades the QLDPC codes decoding performance.
\end{itemize}
These different results confirm the interest of our method of reconciliation based on QTC.
\subsubsection{Secret information}
Secure information is inversely proportional the parameter of eavesdropping attack $s$. $s$ grows up to 1 when Eve succeeds to accedes to all Alice's states. In such a situation, Alice and Bob loss the secret and that is why the protocol should be stopped.

\begin{figure}[h]
\centering
\includegraphics[width=10cm,height=7.5cm]{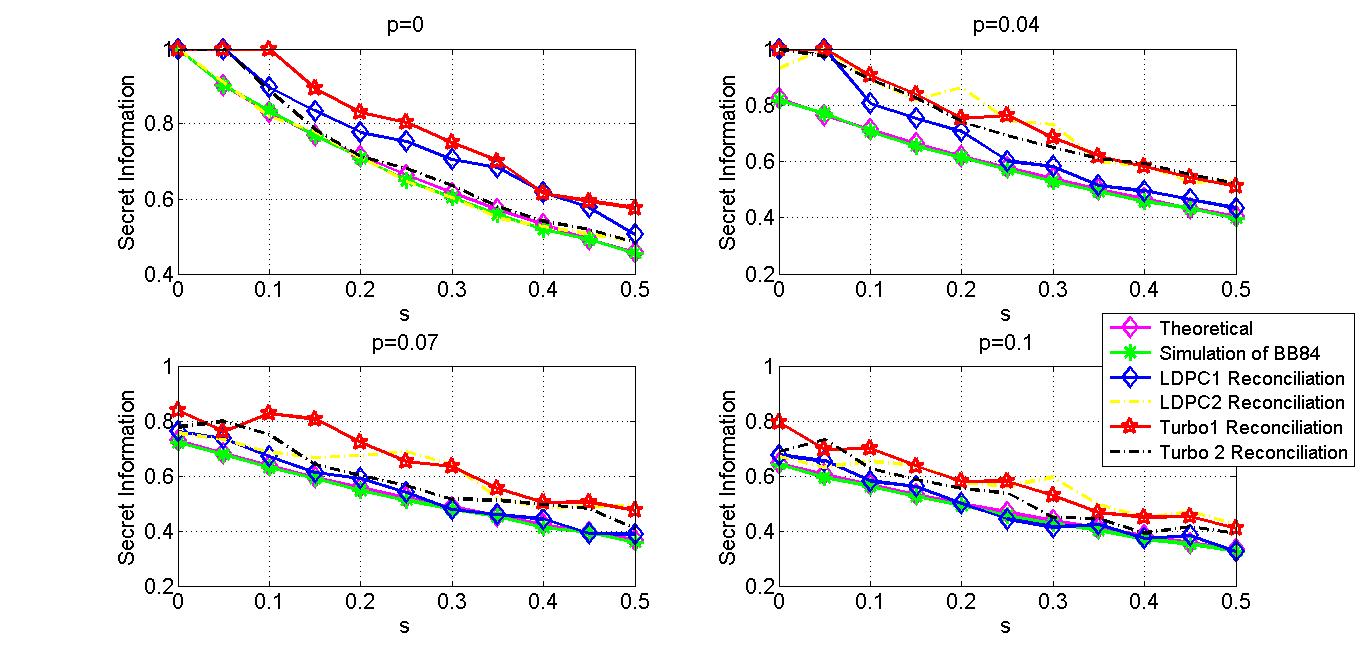}
\caption{Secret information as function of $s$-values for Intercept and Resend attack for different $p$ values.\label{Fig:intercept4}}
\end{figure}

It was established in the literature that we can express the limit of secret information by:
\begin{equation}\label{security}
I_{s}\leq1+\left(1-\delta\right)\log_{2}\left(1-\delta\right)+\delta\log_{2}\left(\delta\right)
\end{equation}
where $\delta$ is the errors's probability.
The Eq.(\ref{security}) allowed us to draw the curves of Fig. \ref{Fig:intercept4}. This figure shows that reconciliation by QTC can improve the security of BB84 protocol better than reconciliation by QLDPC codes.
\subsubsection{Complexity}
The complexity of presented method are evaluated from decoding algorithms. For QLDPC codes algorithm the number of addition and multiplication is linear as a function of the code length $n$. We can therefore consider the complexity as $O (n)$. But for QTC, that complexity can be expressed in a quasilinear manner $O (n log n)$.
In practice, the calculation time is an equally important parameter to complexity for high flow rates and highly interactive diagrams where latency can become a problem. The integration of our method of reconciliation affects little computing time protocol BB84. Both execution times are very close (QTC: $1.58s$ vs QLDPC:$1. 52s$). This confirms once again the advantage of solving the problem of reconciliation by QTC. It is important to note at this stage of our study that these times are those of the execution of our scripts written in C on Intel Core i5 with 2.27 GHz processor and 4GB of RAM.
\subsection{Realization of pratical QKD system}
\begin{figure}[h]
\centering
\includegraphics[width=8cm,height=5cm]{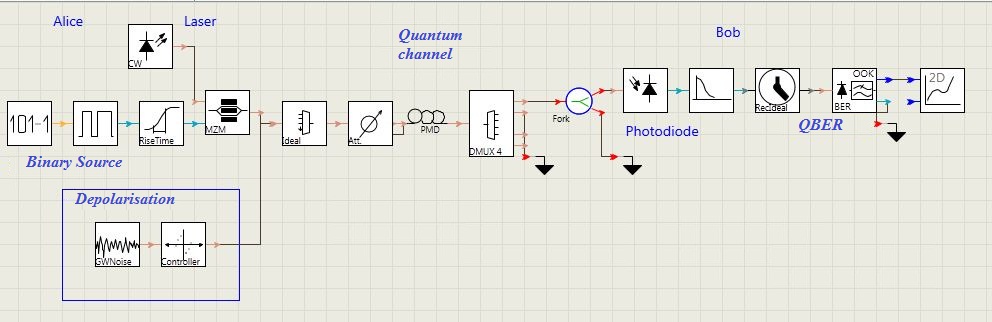}
\caption{Discret quantum cryptography system in the presence of a depolarizing channel.\label{Fig:intercept5}}
\end{figure}

\begin{figure}[h]
\centering
\includegraphics[width=8cm,height=5cm]{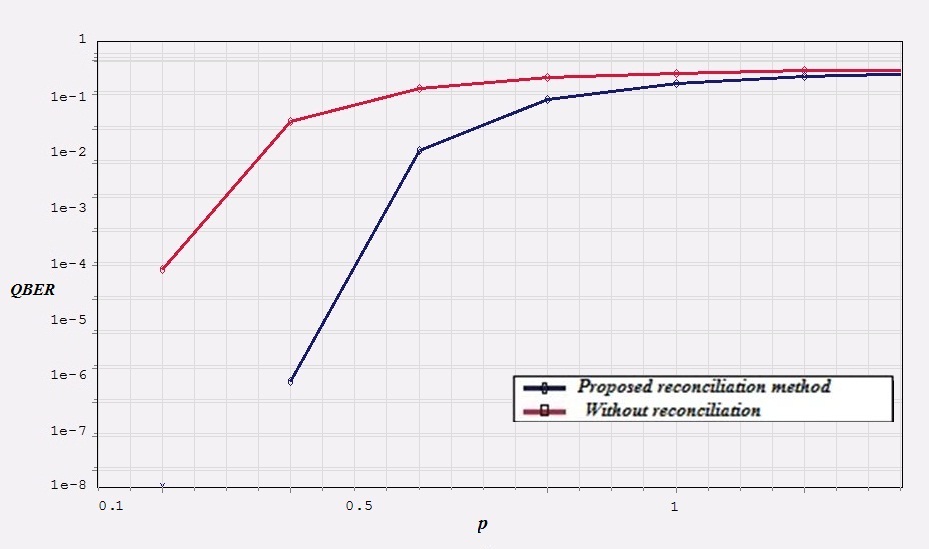}
\caption{QBER as function of $p$-values for $s=0$.\label{Fig:intercept6}}
\end{figure}

We model, design and analyze our proposed QKD systems with VPI Transmission Maker. In our simulations, the method of reconciliation was simulated by our scripts written in C.
On the Alice's side, we work with what is commonly called an external modulation because the optical signal transmitted in the fiber is less troubled by the phenomenon of chirp using this modulation. As shown in Fig.~\ref{Fig:intercept5}, the device is then reduced to a random bit source, a NRZ encoder, a laser source, a Mach-Zehnder modulator (generates relatively low losses of the order of 5dB and does not generate phase shifts), a multiplexer and an attenuator. Quantum regime of the system performance was evaluated by reducing the laser source to obtain an average $\mu = 0.1$ at the output of the transmitter.
At Bob's side, the device comprises a demultiplexer, a photodiode (sensitivity $-12.5dBm$), a synchronization module and a low-pass filter to give the final decision or result (see Fig.~\ref{Fig:intercept5}).
\\We see from the figure Fig.~\ref{Fig:intercept6} that the incoporation of our reconciliation method can undoubtedly reduce the error rate. But the results are a little inferior to those found by simulation of the protocol in the previous section. This difference may in part be explained by the fact that detection is not optimal. Indeed, very often, the noise (shot noise, thermal nois...) which disturbs the electrical signal emitted limit detection photodiodes.
\vspace{-6pt}
\section{conclusion}
\vspace{-2pt}
In this paper, we presented practical QKD system based on QTC and "Slepian-Wolf" model. If we summarize the most interesting questions in the reconciliation research, we can say that it is issue search schemes allowing both a reconstruction of the information with a probability of errors QBER which tends to 0, in this sense our method is more efficient than a concurrent method in the literature, and ideally linear time, in this sense our method is reasonable. The novelty of our work is that it does not stop in the simulation process but it proposed a complete quantum device. Our implementation in a optical fiber environment confirms once again the interest of our reconciliation method.

\end{document}